\documentclass[%
 reprint,longbibliography,preprintnumbers,
nofootinbib,
 amsmath,amssymb,
 aps
]{revtex4-1}
\pdfoutput=1
\usepackage{graphicx}
\usepackage[utf8]{inputenc}
\usepackage{flushend}
\usepackage{dcolumn}
\usepackage{bm}
\usepackage{balance}

\usepackage[normalem]{ulem}

\usepackage[colorlinks = true,
            linkcolor = blue,
            urlcolor  = blue,
            citecolor =green,
            anchorcolor = blue]{hyperref}
\usepackage{verbatim}
\usepackage{color,ulem}
\usepackage[english]{babel}

\usepackage[utf8]{inputenc}
\input Starburst.fd
\newcommand*\initfamily{\usefont{U}{Starburst}{xl}{n}}\initfamily

\newcommand{\beq}{\begin{eqnarray}}
\newcommand{\eeq}{\end{eqnarray}}
\usepackage{amsmath}
\usepackage{tikz}
\usetikzlibrary{decorations.pathmorphing}
\usetikzlibrary{shapes.misc}
\tikzset{cross/.style={cross out, draw=black, minimum size=8*(#1-\pgflinewidth), inner sep=0pt, outer sep=0pt},
cross/.default={1pt}}
\usetikzlibrary{patterns,math}
\begin{document}

\title{Explaining the specific heat of liquids based on instantaneous normal modes}

\author{\textbf{Matteo Baggioli}$^{1,2}$}%
 \email{b.matteo@sjtu.edu.cn}
\author{\textbf{Alessio Zaccone}$^{3,4}$}%
 \email{alessio.zaccone@unimi.it}
 
 \vspace{1cm}
 
\affiliation{$^{1}$Wilczek Quantum Center, School of Physics and Astronomy, Shanghai Jiao Tong University, Shanghai 200240, China}
\affiliation{$^{2}$Shanghai Research Center for Quantum Sciences, Shanghai 201315.}
\affiliation{$^{3}$Department of Physics ``A. Pontremoli'', University of Milan, via Celoria 16,
20133 Milan, Italy.}
\affiliation{$^{4}$Cavendish Laboratory, University of Cambridge, JJ Thomson
Avenue, CB30HE Cambridge, U.K.}

\begin{abstract}
The successful prediction of the specific heat of solids is a milestone in the kinetic theory of matter, due to Debye (1912). No such success, however, has ever been obtained for the specific heat of liquids, which has remained a mystery for over a century. 
A theory of specific heat of liquids is derived here using a recently proposed analytical form of the vibrational density of states (DOS) of liquids, which takes into account saddle points in the liquid energy landscape via the so-called instantaneous normal modes (INMs),  corresponding to negative eigenvalues (imaginary frequencies) of the Hessian matrix. The theory is able to explain the typical monotonic decrease of specific heat with temperature observed in liquids, in terms of the average INM excitation lifetime decreasing with $T$ (in accordance with Arrehnius law), and provides an excellent \color{black} single-parameter \color{black} fitting to several sets of experimental data for atomic and molecular liquids. It also correlates the height of the liquid energy barrier with the slope of the specific heat in function of temperature in accordance with the available data. 
These findings demonstrate that the specific heat of liquids is controlled by the instantaneous normal modes, i.e. by localized, unstable (exponentially decaying) vibrational excitations, and provide the missing connection between anharmonicity, saddle points in the energy landscape, and the thermodynamics of liquids. 
\end{abstract}

\maketitle

Historically, one of the overarching goals of the kinetic theory has always been the rationalization of the specific heat of matter based on its underlying atomic and molecular structure. Classical thermodynamics, revisited in light of modern molecular physics, explains the specific heat of atomic and molecular gases in terms of the equipartition theorem for the various translational and rotational degrees of freedom of the constituent atoms/molecules: the result is the well known Dulong-Petit law, $C_{v}=3N/2$ (constant with $T$), for a monoatomic gas. 

For condensed matter, things become more interesting and more intertwined with modern physics. The case of solids has been essentially solved by Debye in 1912~\cite{Debye}. In his remarkable paper, Debye correctly counted the contribution of plane waves (acoustic phonons) in the isotropic 3d solid to the internal energy, from which he derived the law $C(T)\sim T^{3}$ valid for insulators at low temperature (this does not account for the electronic contribution in metals which is given by the Sommerfeld theory of electronic heat capacity and yiels a $C(T)\sim T$ contribution). Furthermore, in the same paper, Debye presented the famous result for the density of states of phonons in solids, $g(\omega) \sim \omega^{2}$, obtained from the correct way of summing plane wave contributions in a spherical 3d space, together with the ultraviolet cutoff at the Debye wavevector $\omega_{D}$, consistent with atomic-scale granularity of matter.
The correct counting of normal modes in the spherical shell in $k$-space, that is, the $g(\omega) \sim \omega^{2}$,  was the key step that allowed Debye to arrive at the correct result for the specific heat of solids and it strongly relied on the linear dispersion relation  $\omega= v k$ for acoustic phonons. Furthermore, the Debye theory also recovers, correctly once again, the high-temperature limit which is again the Dulong-Petit law mentioned above.

Therefore, we have satisfactory theories of the specific heat for both gases and solids, in agreement with experimental observations, which can be found in any textbooks of statistical physics or solid-state theory \cite{kittel2004introduction} . In light of these successes for gases and solids, it is thus all the more surprising that 100 years after Debye, no satisfactory theory of the specific heat of liquids is available yet. 
Experimental data show that the specific heat of liquids decreases monotonically with temperature upon going from the glass transition or melting transition temperature to higher temperatures.
This behaviour is puzzling because it is clearly in contrast with what is observed in solids, where the specific heat is an increasing function of $T$, and then plateaus at the Dulong-Petit value.

One reason for this state of matters is that the dynamics of atoms and molecules in liquids is strongly anharmonic, which renders the mathematical problem a strongly nonlinear one and intractable from first-principles. This strong anharmonicity also makes concepts such as normal modes, that proved decisive in the Debye theory of specific heat of solids, of less straightforward applicability in the case of liquids. In other words, the basic assumption of Debye theory, i.e. the presence of linearly dispersing propagating (shear) sound waves at small frequencies, must be abandoned. In this sense, a correct description of the specific heat of liquids at small temperatures is inevitably connected to the identification of the low-energy excitations therein, in analogy with acoustic phonons in solids.

Recent advances in our understanding of the specific heat of liquids include the interstitialcy argument by Granato, which heuristically explains the decaying $C(T)$ of liquids in terms of Arrhenius-type relaxation of ``interstitial'' defects \cite{granato}. Though intuitively appealing and simple, this model is not supported by the existence of point-defects in liquids, since there is no underlying regular lattice in liquids that can provide a topologically meaningful definition of interstitials. 
A different explanation of the decaying specific heat of liquids with temperature was suggested by Wallace on the basis of atomic motions through a vast number of random valleys in the energy landscape~\cite{Wallace}.

More recently, Trachenko and co-workers proposed a theory of specific heat in liquids based on standard acoustic phonons \cite{Bolmatov2012} and the k-gap theory \cite{BAGGIOLI20201}. The theory explains the decaying $C(T)$ in liquids as due to the gradual depletion of transverse acoustic phonons (and their shift to higher and higher frequency/momenta) as the temperature is raised. This approach relies on acoustic phonons, whereas at lower momenta/energies one has to deal with overdamped modes ($\omega=-i/\tau$), whose importance for liquids has been established and demonstrated in a broad literature~\cite{Keyes}. The role of these modes is nevertheless not considered in any of the previous approaches.

Modern theories of the liquid state have attempted to extend the concept of normal modes from solids to liquids, following pioneering ideas and work by Zwanzig~\cite{Zwanzig}. This led to the concept of Instantaneous Normal Modes (INMs), which extends the concept of normal modes to liquids, to include the above mentioned overdamped modes. In short, the locally anharmonic dynamics of atoms in liquids leads to many saddle points in the energy landscape. These saddle points are associated with localized unstable (exponentially decaying) modes, with purely imaginary frequency. The imaginary frequencies correspond to negative eigenvalues of the Hessian matrix of the atomistic system. In simpler terms, the anharmonicity leads to locally unbalanced forces between atoms (which are constantly pushed away from their bonding minima by the thermal fluctuations), which then lead to exponentially decaying motions in time with an Arrhenius-dependent time-scale on $T$, i.e. the INMs:
$e^{i \omega^{*}t} \sim e^{-\Gamma t}$, with $\Gamma \sim e^{-U/k_{B}T}$ and $\omega^{*}$ is purely imaginary, $\omega^{*}=-i \Gamma$.

As shown by many numerical studies over the past decades, the INMs dominate the low-frequency and intermediate-frequency sectors of the DOS of liquids~\cite{Keyes,Stratt,Berne}.
At low-frequency, they coexist with one longitudinal acoustic phonon and one transverse diffusive mode (\textit{momentum-shear} diffusion), whereas, at higher frequencies, transverse acoustic phonons only recently have been shown to exist in liquids and to play a role in their thermodynamics at larger energies (the so-called $k$-gap)~\cite{Trachenko}.
Interestingly, these modes define the regime of applicability of \textit{hydrodynamics} \cite{Baggioli:2020loj}, intended as an effective continuum description of fluids.

In this work, we provide a first-principles theory of the specific heat of liquids, which, for the first time, effectively takes into account the intrinsic anharmonicity of liquid dynamics and the fact that the DOS of liquids (derived analytically in recent work \cite{Baggio_liquid}) is dominated by INMs. 
The theory provides an excellent fitting to experimental data of several liquids and correctly recovers the Dulong-Petit law as its high-temperature limit. 
The results presented here provide a long-sought answer to the century-long question about the specific heat of liquids, more than hundred years after Debye's theory for solids.

As it is customary for specific heat calculations, one starts from the total energy of a collection of excitations. For harmonic solids, these excitations are simple harmonic oscillators with frequencies strictly real; in the case of liquids, the frequencies can be imaginary (as for the INMs). States with imaginary frequencies in quantum mechanics are not at all uncommon~\cite{Zeldovich}, and they arise in nuclear physics -- the Gamow states --, and in particle physics, -- the $W$ and $Z^0$ bosons~\cite{Stuart,Sirlin}. These modes are simply called \textit{resonances}, states with a finite lifetime coming from a large imaginary part, which contributes and may even dominate the particle mass and energy~\cite{Zeldovich,Sirlin}. In other contexts, they take the name of \textit{quasinormal modes}; they are intimately linked to non-hermiticity \cite{Bender:2007nj} (i.e. dissipation/relaxation) and experimentally observed even in astronomic black holes collisions \cite{Nollert_1999}.

Here we describe a population of INMs as a weakly-interacting Bose gas, with Hamiltonian given by $\mathcal H = \sum_{q \neq 0}\epsilon_{q}b_{q}^{\dagger}b_{q}$ ~\cite{Khomskii} 
where we do not include the ground state ($T=0$) terms (which is irrelevant since we will later take a derivative with respect to $T$). In the above expression, $b_{q}^{\dagger}$ and $b_{q}$ are the bosonic (Bogolyubov) creation and annihilation operators equipped with standard commutation relations and with associated momentum $q$, while $\epsilon_{q}$ is the energy~\cite{Khomskii}.
We then formally rewrite $\epsilon_{q}\equiv \hbar \omega_{q}$ for the energy of a single boson, where $\omega_{q} \equiv |\omega_{q}|$, as appropriate for unstable bosons~\cite{Zeldovich,Sirlin,Stuart,Keyes}, and we further consider that $b_{q}^{\dagger}b=n_{q}$ where $n_{q}=(e^{\hbar \omega_q/T}-1)^{-1}$ is the Bose-Einstein (BE) occupation number.
Since we have a gauge freedom in defining the ground-state energy (because it obviously does not contribute to the specific heat), we define it as $\hbar\omega_{q}/2$ in order to maintain a formal analogy with the case of solids. 

Hence the energy of a collection of weakly-interacting bosons under the above assumptions can be written as 
\begin{equation}
E=\sum_{q}\frac{\hbar \omega_{q}}{2}\frac{e^{\hbar \omega_q/T}+1}{e^{\hbar \omega_q/T}-1}
\end{equation}
where $q\equiv |\mathbf{q}|$ is the modulus of the momentum, since we are considering isotropic liquids.
In the above, we are working in units such that $k_{B}=1$. 

Using the standard replacement $\sum_{q} \rightarrow \int \frac{d^{3}q}{(2\pi)^3}$, and further introducing the vibrational density of states, $g(\omega)$, defined via $\frac{d^{3}q}{(2\pi)^3} = g(\omega)d\omega$, we arrive at the following integral (which can be found in textbooks) for the specific heat~\cite{Khomskii}:
\begin{equation}
C_{V}(T) = 3N\int_0^\infty \left(\frac{\omega}{2 T}\right)^{2}\sinh{\left(\frac{\omega}{2T}\right)}^{-2}g(\omega)\,d\omega
\label{integral}
\end{equation}
where we have also set $\hbar =k_B=1$.

Upon inserting the normalized Debye DOS, $g(\omega) = 3\omega^{2}/\omega_{D}^{2}$ in the above integral, one readily recovers the low-$T$ limit of the specific heat as $C_{V} \sim T^{3}$, and the high-temperature limit as the Dulong-Petit law, $C_{V} \sim 3N$ (in units of $k_{B}=1$).
\begin{figure}[ht]
\centering
\includegraphics[width=0.45\linewidth]{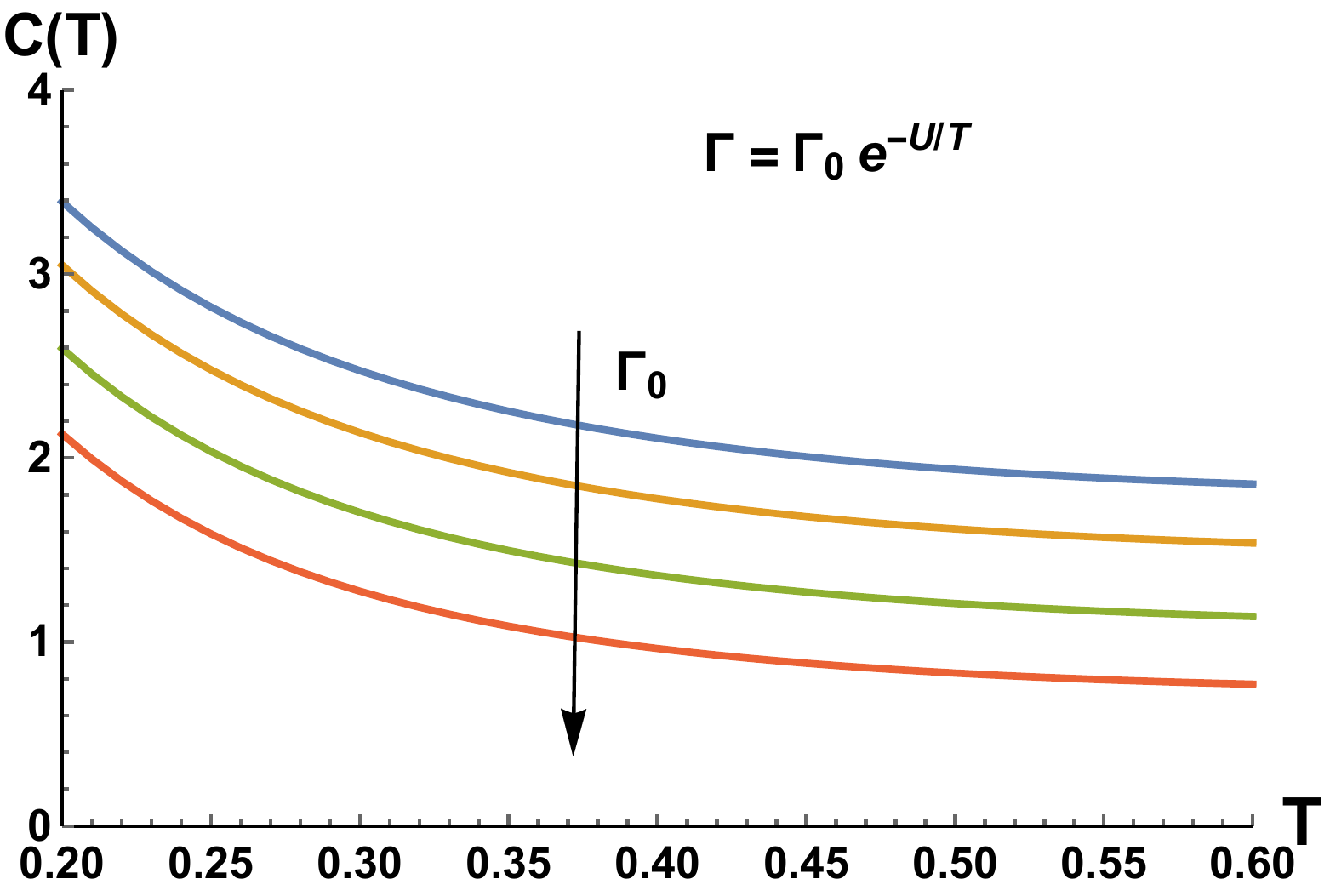}\quad \quad 
\includegraphics[width=0.45\linewidth]{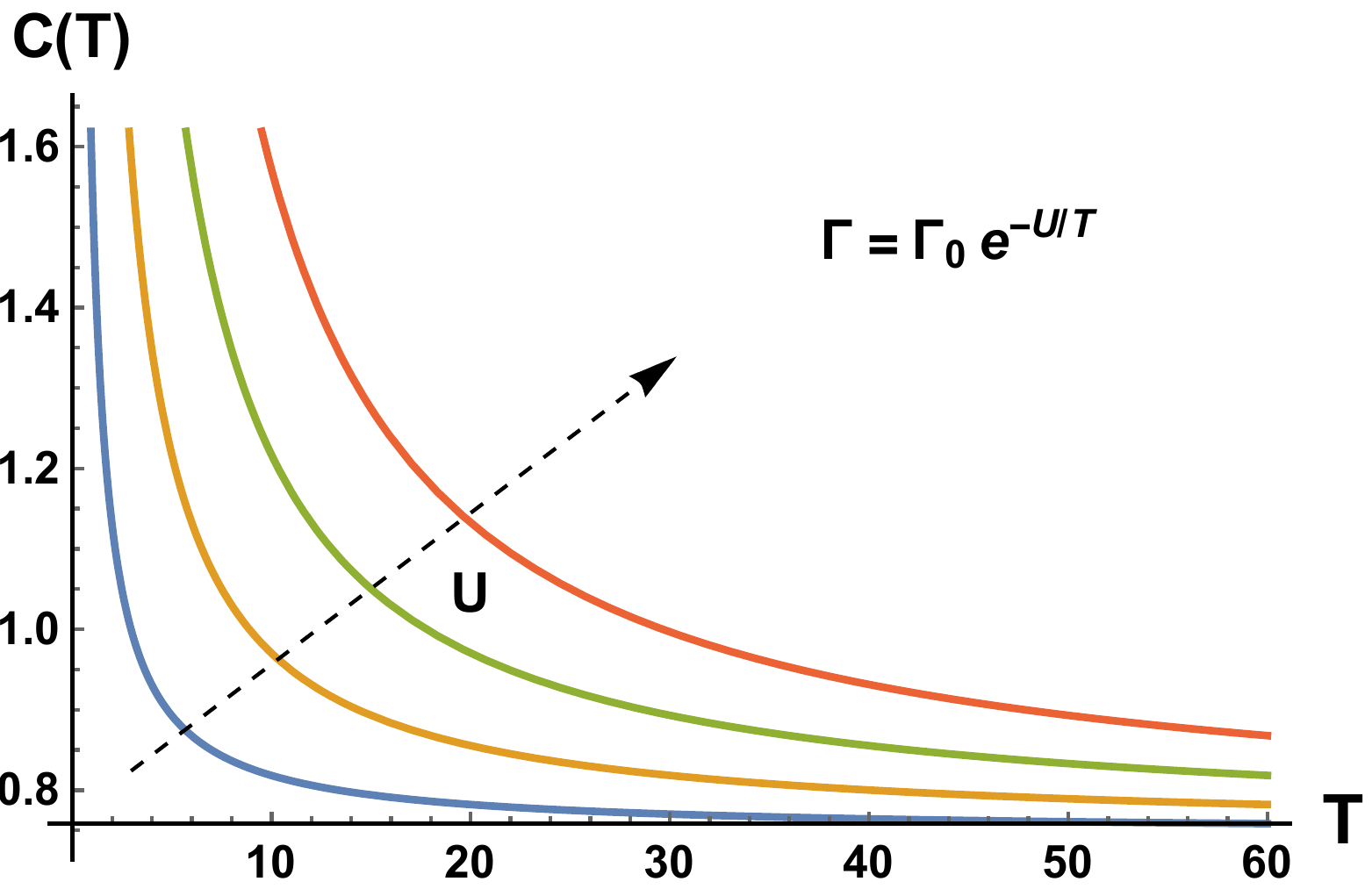}
\caption{The (schematic) theoretical predictions of the model. \textbf{Left: }the dependence of the liquids specific heat on the amplitude of the INMs relaxation rate $\Gamma$. \textbf{Right: } The dependence on the characteristic potential height $U$ for relaxation.}
\label{fig:theory}
\end{figure}\\
Let us now turn to the case of liquids. The starting point is an overdamped equation of motion for particle dynamics,
\begin{equation}
\frac{d\mathbf{v}}{dt}=-\Gamma \,\mathbf{v}, \qquad \text{with}\qquad \Gamma\equiv 1/\tau\,,
\end{equation}
where $\tau$ is the relaxation time and $\Gamma$ is a damping coefficient (the relaxation rate), which for strongly anharmonic excitations represents the (short) lifetime of the excitation.
Taking advantage of a generalization of the Plemelj identity to arbitrary integration pathways in the complex plane, recently it has been possible to derive an analytical form for the DOS of liquids that takes INMs into account \cite{Baggio_liquid}. The final expression has the following form (modulo a normalization factor to ensure that $\int g(\omega)d\omega =1$):
\begin{equation}
g_{\textit{liq}}(\omega)\sim \frac{\omega}{\omega^2+\Gamma^2}\,e^{-\omega^2/\omega_D^2}\,,
\label{DOS}
\end{equation}
where $\Gamma$ is the characteristic relaxation rate of an INM, which exhibits a typical Arrhenius dependence on temperature~\cite{Berne}
\begin{equation}
    \Gamma(T)\,=\,\Gamma_0\,e^{-U/T}.
    \label{Gamma}
\end{equation}
Furthermore, the factor $e^{-\omega^2/\omega_D^2}$ is just a Gaussian cut-off which implements the ``granularity'' of matter at the atomic/molecular scale in terms of the ultraviolet cutoff $\omega_{D}$ and was already introduced in Ref.~\cite{Berne}. We have checked that the main results do not depend essentially on the specific form of the cutoff.

The above Eq.~\eqref{DOS} has been shown in recent work~\cite{Baggio_liquid} to provide an excellent fitting of numerical data of the DOS of Lennard-Jones systems obtained from molecular dynamics simulations in the literature~\cite{Douglas,Berne}.

These formulae, Eqs.~ \eqref{DOS}-\eqref{Gamma}, provide a direct connection between relaxation and vibration in liquids, and play a decisive role in the following description of the specific heat.\\
Upon inserting a normalized form of \eqref{DOS} in \eqref{integral}, it is immediately verified that the limit $T \rightarrow \infty$ of the integral leads $C_{V}=3N$, i.e. the Dulong-Petit law.

We now turn to the dimensional form of the specific heat integral \eqref{integral}
\begin{equation}
C_{V}(T)=k_B\,\int_0^\infty \left(\frac{\hbar\,\omega}{2k_BT}\right)^2\,\sinh \left(\frac{\hbar\omega}{2k_BT}\right)^{-2} g(\omega,T)d\omega
\label{dimensional}
\end{equation}
where $g(\omega,T)$ is given by \eqref{DOS} together with \eqref{Gamma}.

In \eqref{DOS}, acoustic phonons are not explicitly taken into account, because it has been shown in previous work that they are not crucial to reproduce numerical data of DOS of Lennard-Jones liquids~\cite{Baggio_liquid}. 
It is also important to note that, at $T<\Theta_D$ where $\Theta_D$ is the Debye temperature, the BE-related factor $\sinh \left(\frac{\hbar\,\omega}{2\,k_B\,T}\right)^{-2}$ in the integral for the specific heat effectively gives a very low weight to all high-$\omega$ (phonon-type) excitations, whereas it gives a large weight to low-frequency excitations such as the INMs. More precisely, high frequencies could eventually be important only at extremely high temperatures and they cannot possibly be responsible for the low-temperature (above melting transition) decay typical of liquids. Indeed, as we will prove, there is no need to take into account high frequency modes (e.g. emerging shear waves in the k-gap model \cite{BAGGIOLI20201}) to reproduce the experimental trends.

Furthermore, quoting from Born and Huang~\cite{Born-Huang}, at $T>\Theta_D$, the specific heat is not sensitive to the specifics of the frequency distributions and the Einstein model provides a correct estimate in terms of high-energy atomic/molecular vibrations with $\omega \sim \omega_{D}$ or larger (intramolecular vibrations). Hence, in this high-temperature regime, phonons, as collective lattice vibrations, do not exist anymore, while the high-frequency non-collective (gas-like) vibrations contribute a constant (independent of $T$) to the specific heat~\cite{Born-Huang}.
These arguments suggest that the influence of the INMs on the specific heat and on its observed decay with $T$ could possibly be the dominant one.

Illustrative calculations of the specific heat using the above theory are shown in Fig.\ref{fig:theory}. It is clear from these theoretical calculations that the temperature dependence of the specific heat is mostly controlled by the relaxation rate of excitation lifetime $\Gamma$ and its Arrhenius dependence on $T$. In particular, despite the dimensionful pre-factor $\Gamma_0$ produces only a vertical shift in the $C(T)$ function (left panel of Fig.\ref{fig:theory}), the energy barrier $U$ plays a much more fundamental role. It determines the curvature of the specific heat; the larger the potential energy $U$, the slower the temperature decay of the specific heat (right panel of Fig.\ref{fig:theory}).

This Arrhenius dependence was fortuitously captured by Granato's interstitial defect argument, although its true physical origin resides in the INMs and in the many saddle points of the energy landscape.
From a physical point of view, the decay of $C(T)$ with increasing $T$ is caused by the decrease of the average lifetime of the INM excitations, which is equal to $\Gamma^{-1}$. Hence, since the heat is stored by the INMs,  as the dominant vibrational excitations in liquids, the fact that their lifetime decreases with increasing $T$ leads to a lower capability of storing heat in the vibrational excitations.

This picture is confirmed by the fact that the specific heat is reduced upon increasing the strength of the INMs relaxation rate $\Gamma_0$, i.e. upon decreasing their lifetime. Moreover, the model directly shows that, by increasing the characteristic potential height $U$ of the anharmonic liquid landscape, the specific heat grows. This can be simply explained by the fact that a higher barrier suppresses the probability of the molecular rearrangements responsible for the INMs dynamics and therefore makes their lifetimes longer. This is fully consistent with the emerging picture of heat being stored in the INMs, in liquids.

\begin{table}
 \begin{tabular}{||c||c c c c  c||} 
 \hline
Liquids: &  Xe & Kr & Ne & Ar  & N$_2$\\ [0.5ex] 
 \hline\hline
 $ \omega_D^*$ [K]& 64 & 72.1 &  74.6&  93.1 & 86\\
 \hline
 $ U^*$ [K]& 226.1  & 162.5 & 33.9 & 116.7 & 102.12\\
 \hline
 $ \Gamma_0$ [K]& 240 & 100 & 80 & 60  & 29\\
\hline
\end{tabular}
\caption{The numerical values used in the fitting procedure. The symbol $^*$ indicates that the values are not obtained from the fit but they are fixed with the literature data \cite{PhysRev.142.490,pmid10001114,doi:10.1080/00268976.2016.1246760}. The only free parameter is $\Gamma_0$.}
\label{data}
\end{table}

\begin{figure}[ht]
\centering
\includegraphics[width=0.9\linewidth]{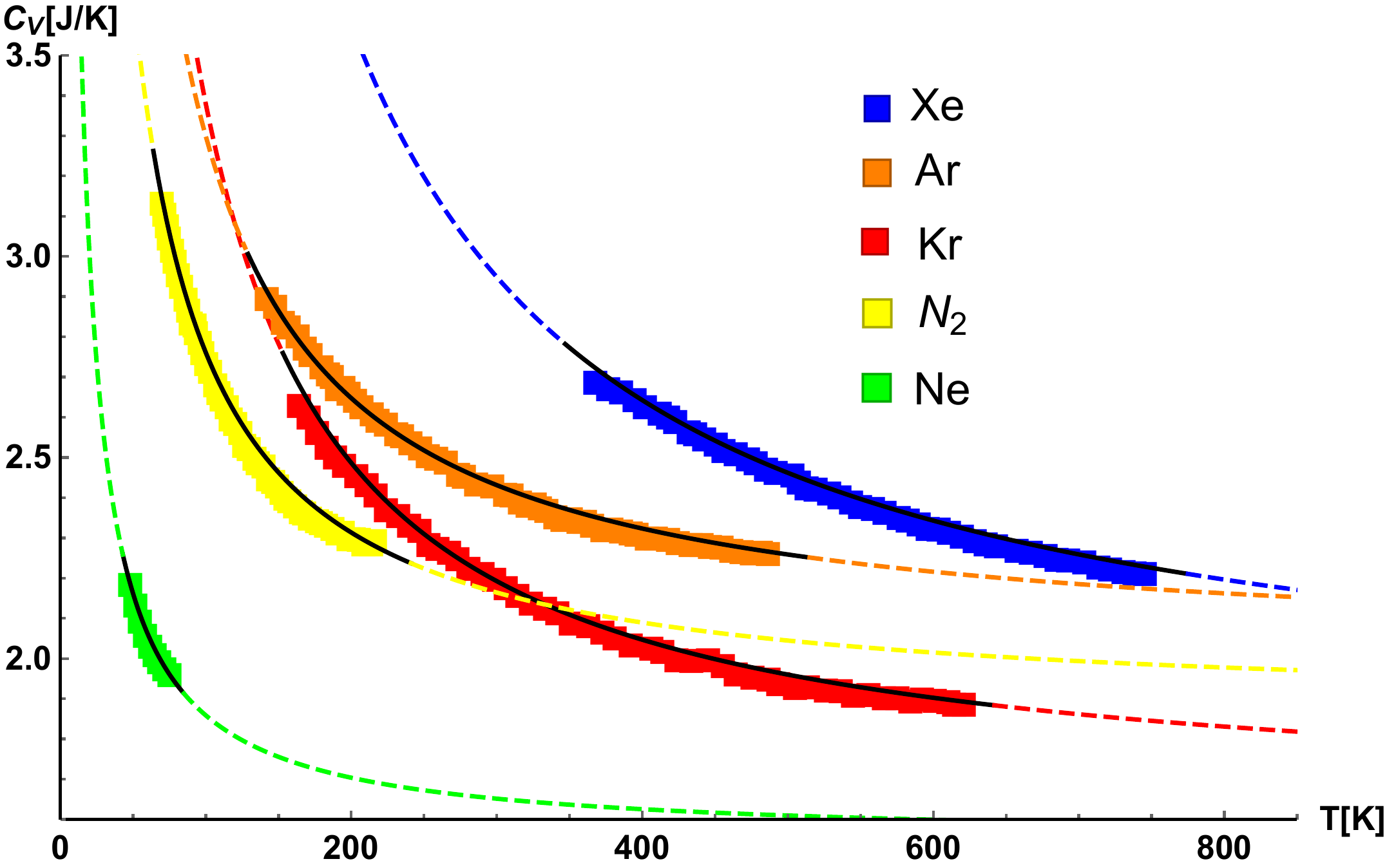}
\caption{The comparison between the model, Eqs.\eqref{DOS}-\eqref{dimensional}, and experimental data for four different liquids. The experimental data are taken from \cite{Wallace,nist}. The value of the various parameters is displayed in Table \ref{data}.}
\label{fig:fit1}
\end{figure}

\color{black}
We now turn to the fitting procedure and the main results of our analysis. Combining Eq.\eqref{DOS} and Eq.\eqref{Gamma}, our model displays three physical parameters: the Debye frequency $\omega_D$, the activation energy $U$ and the relaxation rate prefactor $\Gamma_0$. The first two parameters for simple liquids are well-known and they are fixed to their literature values \cite{PhysRev.142.490,pmid10001114,doi:10.1080/00268976.2016.1246760} displayed in Table \ref{data}. The activation energy is taken to be equivalent to the height of the Lennard-Jones energy barrier $\epsilon$. All in all, our fitting procedure involves a single fitting parameter $\Gamma_0$.
In Fig. \ref{fig:fit1}, we present a series of comparisons between the specific heat calculated using \eqref{DOS} inside
the specific heat integral \eqref{dimensional} and experimental data of simple liquids of various nature, but all reasonably well approximated by the Lennard-Jones potential. The obtained values for the relaxation rate scale $\Gamma_0$ are shown in Table \ref{data}.
In all instances, the fitting is excellent and perfectly captures the decline of the specific heat with increasing temperature, explained by the present theory in terms of reduced lifetime of INMs. The results show, as already anticipated, that, the larger the characteristic energy $U$ (which is related to $\epsilon$), the larger the specific heat and the slower its temperature decay. This confirms once more not only the validity of our theory but also its predictive power able to connect microscopic features, such as the characteristic potential barrier $U$, to macroscopic thermodynamic observables, such as the temperature dependence of the specific heat.

\begin{figure}[ht]
    \centering
    \includegraphics[width=0.9\linewidth]{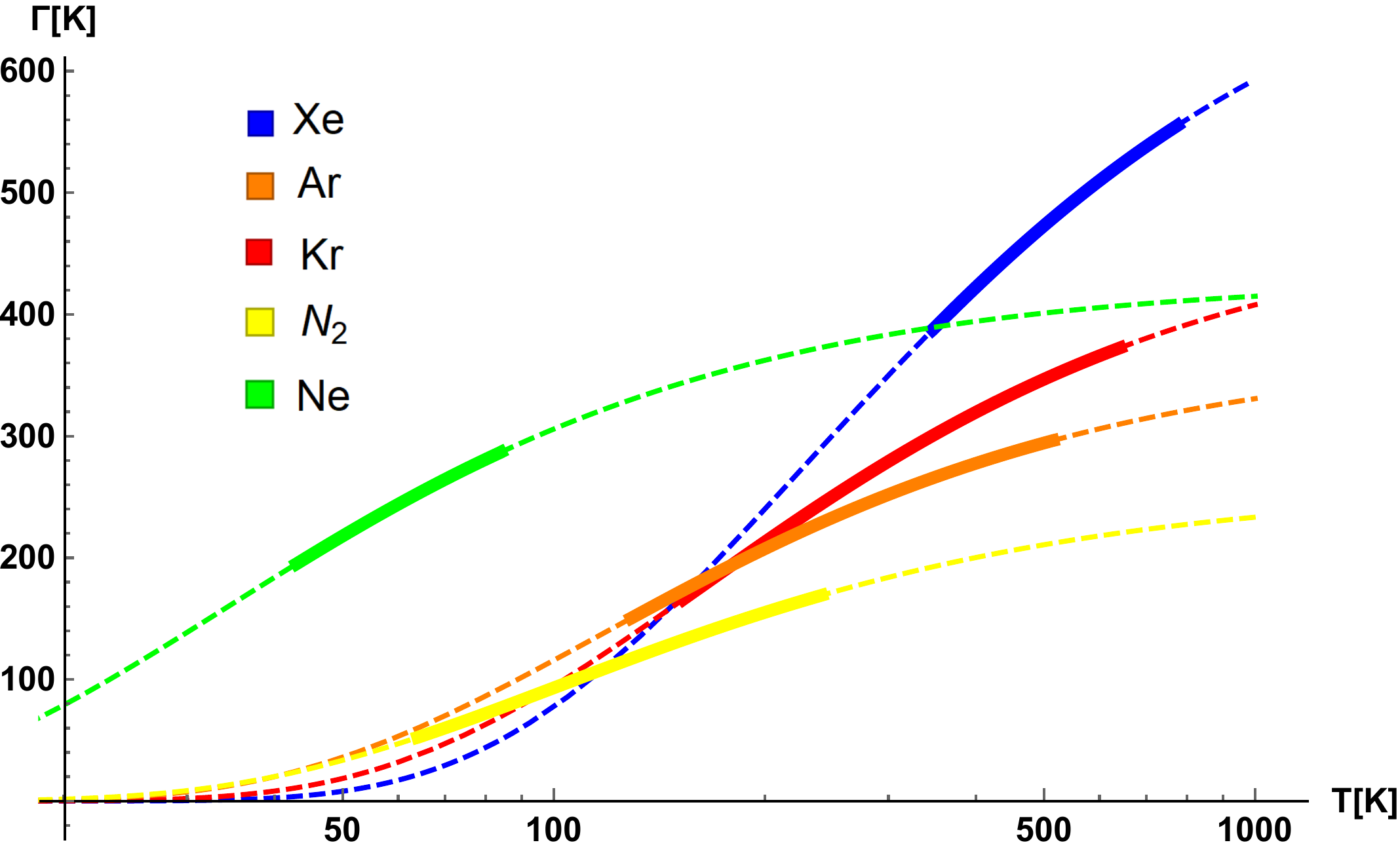}
    \caption{The temperature dependent INM relaxation rate $\Gamma(T)$ obtained by using the single-parameter fitting in Table \ref{data}. The solid portion of the curves is the one corresponding to the temperature range of the experimental data fitted.}
    \label{fig:all}
\end{figure}

In order to emphasize the predictive power of our theory, and the excellent agreement with the data, we re-present the INMs temperature dependent relaxation rate $\Gamma(T)$ using the parameters obtained from the fits in Fig. \ref{fig:all}. For all the liquids analyzed, we find a relaxation rate of the order of $1/$ps. According to transition state theory \cite{RevModPhys.62.251}, the molecular hopping (attempt) rate is directly proportional to the INMs relaxation rate, which corresponds to the (negative) curvature of the potential landscape. Interestingly, our order of magnitude estimate of the single fitting parameter $\Gamma_{0}$, coincides with the values reported in the literature, see for example \cite{Berne}.\color{black}

In summary, the above theory provides a definitive answer to the mystery of liquid specific heat and ideally completes the agenda of the kinetic theory of matter, set over 100 years ago by Debye, Einstein, Planck and co-workers.

As in Debye's work~\cite{Debye} for solids, the crucial step for the successful derivation of the specific heat, also in the case of liquids relies on finding the correct form of the vibrational density of states (DOS). Debye derived his famous $T^{3}$ law for the specific heat of solids by correctly counting 3d plane waves in an isotropic solid, leading to the Debye vibrational density of states, $\sim \omega^{2}$. 
Here we did the same for liquids, where the relevant excitations are not plane waves/phonons but the instantaneous normal modes (INMs), i.e. overdamped relaxations from saddle points in the energy landscape. This leads to a DOS for liquids $\sim \omega$ at low frequency \cite{Baggio_liquid}, whose form is given in Eq.\eqref{DOS}. In turn, this DOS leads to a monotonically decreasing $C(T)$ with increasing $T$, as a result of Arrhenius-type relaxation of INMs, and recovers the Dulong-Petit plateau in the high-T limit. 

These results fill the gap in our understanding of thermal and vibrational properties of condensed matter.

Finally, given the success of the theory by Trachenko and co-workers \cite{Bolmatov2012}, it is important to draw some comparisons. Given our results, it is clear that the key point in their treatment is not the presence of propagating shear waves, which appear at large momenta and frequencies (at least at momenta larger then $\sqrt{2}k_g$), but rather the collection of overdamped modes below that point. In particular, the k-gap dispersion relation \cite{BAGGIOLI20201} displays purely relaxing modes below $k=k_g$. Not only that, but even between $k_g<k<\sqrt{2}k_g$, the acoustic waves are mostly overdamped, and therefore more similar in nature to INMs than to propagating shear waves. Moreover, our results are in agreement with those of Ref.~\cite{PhysRevLett.125.125501} where the heat capacity decreases by increasing the k-gap momentum. Indeed $k_g \sim \Gamma$; a larger k-gap implies a shorter lifetime for the relaxational modes $\omega=-i \Gamma$ and therefore a lower specific heat as explained by our theory.

Following the ideas of \cite{Baggioli:2021ntj}, it would definitely be interesting to achieve a more fundamental understanding of this relaxation time scale based on symmetries rather than microscopic mechanisms, in analogy to the modern formulation of phonons and Debye theory in terms of the spontaneous symmetry breaking of spacetime translations.

\subsection*{Acknowledgments} 
We thank K.Trachenko for fruitful discussions and useful comments.
A.Z. acknowledges financial support from US Army Research Office, contract nr. W911NF-19-2-0055. M.B. acknowledges the support of the Shanghai Municipal Science and Technology Major Project (Grant No.2019SHZDZX01) and
of the Spanish MINECO “Centro de Excelencia Severo Ochoa” Programme under grant
SEV-2012-0249.
\bibliographystyle{apsrev4-1}

\bibliography{sample}

\end{document}